\documentclass[aps,prl,reprint,superscriptaddress]{revtex4-2}

\usepackage{graphicx}
\usepackage{todonotes}
\usepackage{amsmath}
\usepackage{soul}

\begin{document}

\title{Enhancing Photon Indistinguishability of Spectrally Mismatched Single Photons by Cavity Floquet Engineering}

\author{Jia-Wang Yu}
\affiliation{College of Information Science and Electronic Engineering, Zhejiang University, Hangzhou 310027, China}
\affiliation{State Key Laboratory of Silicon and Advanced Semiconductor Materials, ZJU-Hangzhou Global Scientific and Technological Innovation Center, Zhejiang University, Hangzhou 311200, China}
\author{Xiao-Qing Zhou}
\affiliation{Department of Physics, School of Science, Westlake University, Hangzhou 310030, China}
\affiliation{Institute of Natural Sciences, Westlake Institute for Advanced Study, Hangzhou 310024, China}
\author{Zhi-Bo Ni}
\affiliation{College of Information Science and Electronic Engineering, Zhejiang University, Hangzhou 310027, China}
\affiliation{State Key Laboratory of Silicon and Advanced Semiconductor Materials, ZJU-Hangzhou Global Scientific and Technological Innovation Center, Zhejiang University, Hangzhou 311200, China}
\author{Xiao-Tian Cheng}
\affiliation{College of Information Science and Electronic Engineering, Zhejiang University, Hangzhou 310027, China}
\affiliation{State Key Laboratory of Silicon and Advanced Semiconductor Materials, ZJU-Hangzhou Global Scientific and Technological Innovation Center, Zhejiang University, Hangzhou 311200, China}
\author{Yi Zhao}
\affiliation{College of Information Science and Electronic Engineering, Zhejiang University, Hangzhou 310027, China}
\affiliation{State Key Laboratory of Silicon and Advanced Semiconductor Materials, ZJU-Hangzhou Global Scientific and Technological Innovation Center, Zhejiang University, Hangzhou 311200, China}
\author{Hui-Hui Zhu}
\affiliation{College of Information Science and Electronic Engineering, Zhejiang University, Hangzhou 310027, China}
\affiliation{State Key Laboratory of Silicon and Advanced Semiconductor Materials, ZJU-Hangzhou Global Scientific and Technological Innovation Center, Zhejiang University, Hangzhou 311200, China}
\affiliation{College of Integrated Circuits, Zhejiang University, Hangzhou 311200, China}
\author{Chen-Hui Li}
\affiliation{College of Information Science and Electronic Engineering, Zhejiang University, Hangzhou 310027, China}
\affiliation{State Key Laboratory of Silicon and Advanced Semiconductor Materials, ZJU-Hangzhou Global Scientific and Technological Innovation Center, Zhejiang University, Hangzhou 311200, China}
\affiliation{Research Institute of Intelligent Computing, Zhejiang Lab, Hangzhou 311121, China}
\author{Feng Liu}
\affiliation{College of Information Science and Electronic Engineering, Zhejiang University, Hangzhou 310027, China}
\author{Chao-Yuan Jin}
\email{jincy@zju.edu.cn}
\affiliation{College of Information Science and Electronic Engineering, Zhejiang University, Hangzhou 310027, China}
\affiliation{State Key Laboratory of Silicon and Advanced Semiconductor Materials, ZJU-Hangzhou Global Scientific and Technological Innovation Center, Zhejiang University, Hangzhou 311200, China}
\affiliation{College of Integrated Circuits, Zhejiang University, Hangzhou 311200, China}

\date{\today}

\begin{abstract}
We theoretically propose a scheme to enhance the photon indistinguishability of spectrally mismatched single photons via Floquet-engineered optical frequency combs (OFCs) in cavity quantum electrodynamic systems. By periodically modulating two distinct single-photon states under a modulation frequency which is exactly equal to the spectral mismatch of two cavity modes, a pair of single-photon frequency-comb (SPFC) states is prepared energy-conservatively based on full unitary operations. The two states show high indistinguishability with an ideal $g^{(2)}_\mathrm{HOM}(0)$ down to zero due to the superposition of intensity-matched single-photon states coherently distributed across the teeth of the combs.
\end{abstract}

\maketitle

Perfectly indistinguishable single photons across all degrees of freedom are crucial for key applications such as quantum communication and quantum computation \cite{horodecki2009quantum, fan2016integrated, elshaari2020hybrid}. However, photons generated from different sources, whether from heterogeneous or homogeneous devices, usually exhibit distinct profiles in their wave packets that compromise photon indistinguishability. The spectral mismatch of two single-photon states, even in very small quantities, poses significant obstacles to the scalability of photonic quantum networks and the integration of single-photon sources into large-scale quantum information processing (QIP) systems \cite{aharonovich2016solid, castelletto2020silicon}.

Traditionally, frequency conversion techniques such as sum-frequency generation \cite{takesue2008erasing} and spectral shearing \cite{wright2017spectral, zhu2022spectral} are employed to bridge the frequency difference and improve the indistinguishability of single photons. These frequency-shifting techniques are usually modeled as a unitary operation in photon-number space, which requires a minimum energy cost according to recently developed no-go theorems \cite{tajima2018uncertainty, chiribella2021fundamental}. Fundamental constraints dictate that the energy cost increases as the spectral mismatch widens, making these frequency-shifting techniques costly options under significant spectral mismatches \cite{yang2022energy}. 

Here, we propose a scheme based on the tailoring of single-photon spectra with cavity Floquet engineering, which is a newly emerging tool for manipulating quantum systems through time-periodic modulation of Hamiltonians \cite{iorsh2022floquet,zhou2023pseudospin,tong2025observation}. As a result, this energy-conserving operation is not constrained by the fundamental limitation, and the driving energy cost does not exhibit a direct dependence on the spectral mismatch, thereby enabling fully unitary operation even at large spectral mismatches. For instance, when the spectral mismatch is as large as 50 GHz and the photon duration is 500 ps, the driving energy cost of our scheme is orders of magnitude lower than that of the spectral shearing technique. The underlying mechanism for this energy cost difference can be fully explained within the framework of quantum thermodynamics. The proposed scheme based on cavity-enhanced modulation offers the potential for a further reduction in energy demand \cite{zhou2024cavity, jin2014switch}.

\begin{figure}[h]
\centering
\includegraphics[width=0.4\textwidth]{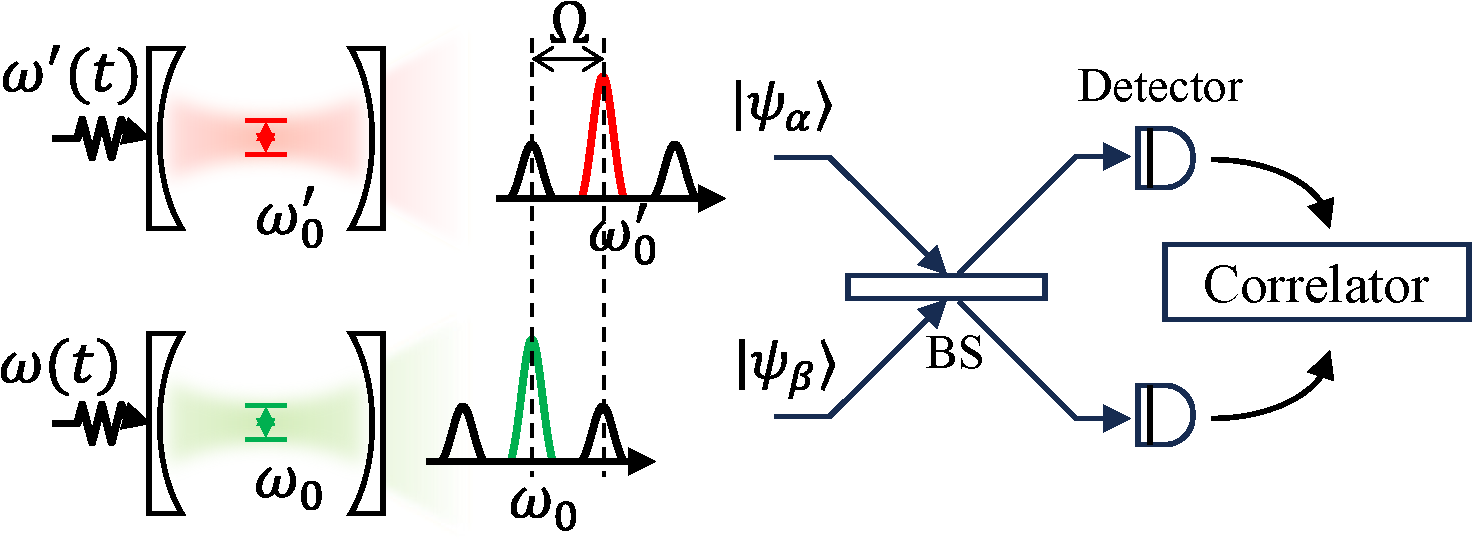}%
\caption{A schematic for the photon-indistinguishability enhancement based on cavity Floquet engineering. Two SPFC states are prepared by modulating two spectrally mismatched single photon states, which enlarges the spectral overlap between two states and thereby enhances photon indistinguishability. The $g^{(2)}_\mathrm{HOM}(0)$ is then measured for two SPFCs with Hong–Ou–Mandel (HOM) setup.\label{fig:scheme}}
\end{figure}

Our photon-indistinguishability-enhancement scheme based on the modulation of cavity-mode frequency is illustrated in Fig. \ref{fig:scheme}. Two sets of distinct single photons were originally generated from two detuned optical cavities at frequencies $\omega_0$ and $\omega_0'$ with a frequency detuning of $\Omega=\omega_0'-\omega_0$. The two cavities are modulated separately by applying periodic signals $\omega(t)$ and $\omega'(t)$ on the cavity reflectors. This modulation process generates single-photon frequency-comb (SPFC) states~\cite{liao2016single,ren2014simultaneous}, which are widely used in spectral QIP encoding schemes~\cite{lu2019quantum, lukens2016frequency, yamazaki2023linear}. The SPFC state, where one single photon is coherently distributed across multiple and equally spaced frequency components, is written as $|\psi_\alpha\rangle=\sum_n{s_{\alpha,n}}|\omega_0+n\Omega\rangle$ where $|\omega_0+n\Omega\rangle$ is a single-photon state at frequency $\omega_0+n\Omega$. The generation process can be described by a full unitary transformation in both the photon-number and energy space. The photon-number expectations of all components are preserved at one, whereas the central frequency of SPFC states is not shifting to preserve the total energy \cite{liao2016single,ren2014simultaneous}. In the meantime, the spectral overlap between two SPFCs can be enhanced by optimizing the modulation function, thereby facilitating photon indistinguishability.

Photon indistinguishability is quantified by the coincidence probability in a Hong–Ou–Mandel (HOM) experiment, described by the second-order coherence function \cite{fischer2016dynamical},
\begin{align}
    |\psi_\alpha\rangle&=U_\alpha|\omega_0\rangle,\nonumber \\
    |\psi_\beta\rangle&=U_\beta|\omega_0+\Omega\rangle,\nonumber \\
    g^{(2)}_\mathrm{HOM}(\tau)&=\frac{1}{2}\left[1-\left| \langle\psi_\alpha |D(\tau)|\psi_\beta\rangle \right|^2\right]\label{eq:g2}, 
\end{align}
where $D(\tau)$ denotes the temporal delay of $\tau$ for the wave function of single photons, and $U_{\alpha,\beta }$ represent the modulation operations. These operators can be expressed as $U_\alpha=\sum_{k,n}s_{k-n}|\omega_0+k\Omega\rangle\langle \omega_0+n\Omega|$, where the coefficients $s_k$ are determined by the modulation signal $s(t)$, as $s_k=\int_0^T e^{is(t)}e^{ik\Omega t}dt /T$ \cite{lukens2016frequency}. So is $U_\beta$ given by $U_\beta=\sum_{k,n}s'_{k-n}|\omega_0+k\Omega\rangle\langle\omega_0+n\Omega|$ with the coefficient $s'_k=\int_0^T e^{is'(t)}e^{ik\Omega t}dt /T$.  Our objective is to minimize the second-order coherence at zero delay $g^{(2)}_\mathrm{HOM}(0)=\left[1-|\langle\omega_0| U^\dagger_\alpha U_\beta |\omega_0+\Omega\rangle|^2\right]/2$, which reaches its minimum when the coefficients satisfy $s_n = e^{i\phi}s'_{n-1}$ \cite{supp}. Here, $\phi$ is a real number that represents the phase difference between $s_n$ and $s'_{n-1}$.

Various modulation strategies can satisfy such a condition that generating the required SPFCs. In the following discussion, a simple approach is taken in which both photons undergo identical modulation, i.e., $U_\alpha=U_\beta$. The condition reduces to 
\begin{equation}\label{eq:cond}
    s_n=e^{i\phi}s_{n-1},
\end{equation}
which indicates a flat-top SPFCs with constant phase difference between adjacent comb teeth. An established method for generating flat-top SPFCs is to use a periodic parabolic modulation \cite{torres2008lossless,kolner1989temporal}, often referred to as a "time lens" \cite{wei2017unveiling}. The parabolic modulation signal reads $s(t)=\frac{A\Omega^2}{\pi^2}(t^2-\frac{2\pi}{\Omega}t)$, where $\Omega$ denotes the modulation frequency, and $A$ is preserved as a dimensionless modulation depth. 
\begin{figure}[h]
    \centering
    \includegraphics[width=0.45\textwidth]{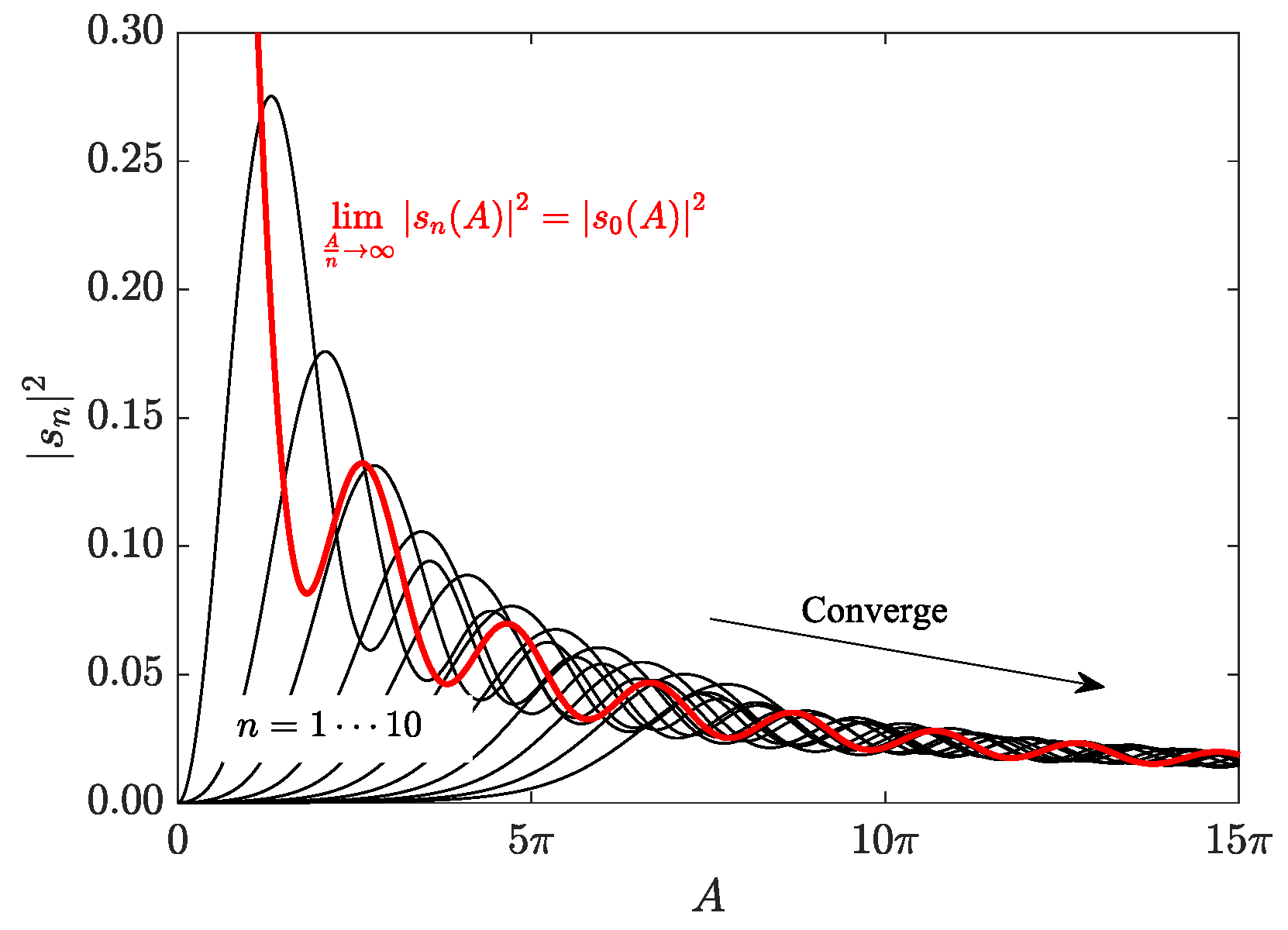}%
    \caption{\label{fig:inten} Intensity of SPFC sidebands as a function of modulation depth $A$. For visual guidance, The thick red line represents the sideband intensity when $A \gg n$. The intensity of sidebands converge generally toward the red line with high modulation depth, resulting in a uniform intensity distribution.}
\end{figure}
Fig. \ref{fig:inten} shows the intensity of the SPFC sideband $|s_n|^2$ as a function of $A$. Since $|s_{-n}|=|s_n|$, we will only discuss the sidebands with $n>0$ \cite{supp}. Once $A$ becomes much larger than $n$, the sideband intensity begins to oscillate and gradually converge to
\begin{equation}\label{eq:snll}
    \lim_{\frac{A}{n} \to \infty}s_n (A) = -\frac{\sqrt{i\pi}e^{-i (A-n\pi)}}{2\sqrt{A}} \text{erf}\left( i\sqrt{iA} \right) ,
\end{equation}
where $\text{erf}(z)$ represents the Gaussian error function. This approximation reveals that the SPFC is flat-top and phase-locked with a high modulation depth $A$, which eventually results in the satisfaction of the condition Eq. (\ref{eq:cond}) and the convergence of $g^{(2)}_\mathrm{HOM}(0)$ toward 0. 

In practice, a moderate modulation depth $A$ can compromise the intensity uniformity of the sidebands and the phase-locking conditions in between. The problem can be alleviated by introducing a time delay $\delta$ between the two SPFCs, an experimentally straightforward way to adjust the relative phase of the sidebands. When a time delay of $\delta$ is applied to the state $|\psi_\beta\rangle$, i.e. $D(\delta)|\psi_\beta\rangle=\sum_n s_{n-1}e^{-i(\omega_0 + n\Omega) \delta}|\omega_0+n\Omega\rangle$, the phase difference of two SPFCs at each frequency $\omega_0+n\Omega$ is now $\Delta \phi_n =\mathrm{arg}(s_n)-\mathrm{arg}(s_{n-1}e^{-in\Omega\delta})$. The overlap between two states is
\begin{equation}\label{eq:phase}
    \left|\langle \psi_\alpha| D(\delta) | \psi_\beta\rangle\right| = \left|\sum_n e^{-i\Delta\phi_n}  s^*_n s_{n-1} \right|,
\end{equation}
which reaches its maximum at a time delay of $\delta_{opt}=\frac{\pi^2}{2A\Omega}$ (see detailed analysis in Supplementary \cite{supp}), as shown in  Fig. \ref{fig:phase}(b). Hence, optimal phase locking $(\Delta\phi_n\approx \phi)$ is evidenced by the minimum variation of the phase difference $\Delta\phi_n$, shown in Fig. \ref{fig:phase}(a) along the red dashed line. 

\begin{figure}[h]
    \centering
    \includegraphics[width=0.45\textwidth]{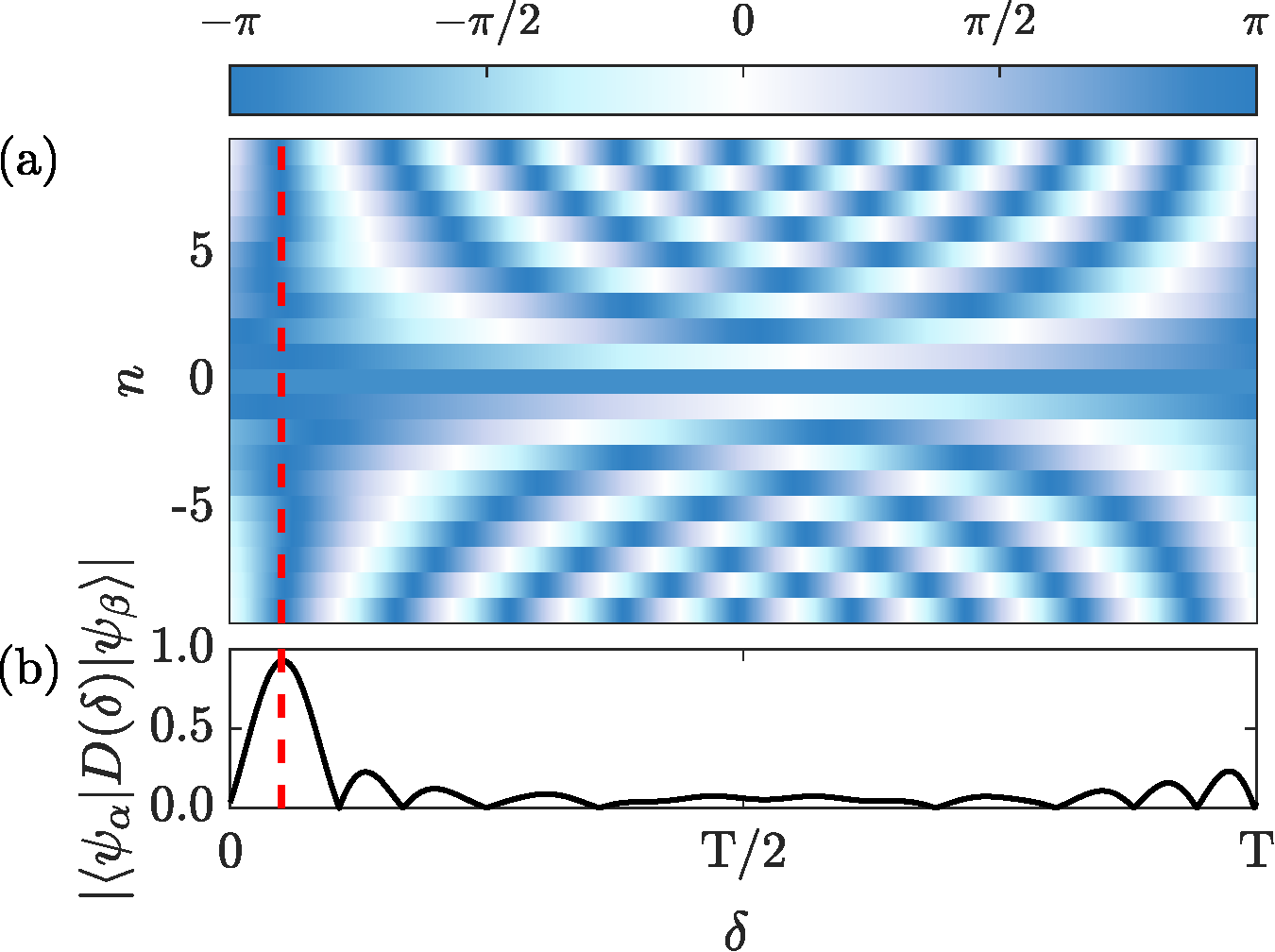}%
    \caption{\label{fig:phase}(a) The phase difference $\Delta \phi_n$  of two SPFCs at the frequency $\omega_0+n\Omega$. The red dashed line extended from (b) marks the optimized phase-locking condition, i.e. minimum variation in $\Delta \phi_n$ at a time delay of $\delta_{opt}=\frac{\pi^2}{2A\Omega}$. (b) The wave-function overlap between SPFC states $|\langle \psi_\alpha|D(\delta)|\psi_\beta\rangle|$ as a function of the signal delay $\delta$. The red dashed line indicates the time delay for a maximum overlap corresponding to the optimal phase-locking condition shown in (a). These figures demonstrate that achieving phase locking through adjusting the time delay $\delta$ (as shown in (a)) directly leads to the maximization of the overlap between the SPFC states (as shown in (b)). The calculations use parameters of $A=5\pi$ and $\Omega=50/(2\pi)$ GHz.}
\end{figure}

To validate this approach, we employ a two-level system (TLS) coupled to the modulated cavity field as a platform to generate flat-top SPFCs, providing a theoretical framework for approximating quantum frequency modulation processes~\cite{silveri2017quantum}. Experimentally, such a modulated cavity scheme can be realized through semiconductor and atomic systems \cite{anguiano2017micropillar, weiss2016surface, tian2022all, jin2014ultrafast, jin2009vertical}. The coupled system between the TLS and modulated cavity forms a unit device, whose Hamiltonian is given by \cite{janowicz1998evolution,yang2004interactions,law1995modification, law1994effective}
\begin{equation}
    H(t)=\left[ \omega_c+\omega_m(t) \right] a^\dagger a + \omega_c \sigma^\dagger \sigma + g(\sigma^\dagger a + a^\dagger \sigma),
\end{equation}
where $a^\dagger(a)$, $\sigma^\dagger(\sigma)$ are the creation (annihilation) operators for the cavity field and TLS. The coupling rate between the TLS and the cavity is $g$, the dissipation rates of these two are $\gamma$ and $\kappa$, respectively. The time delay mentioned above is further introduced by moving the modulation signal $\omega(t)=\omega_m(t)$ and $\omega'(t)=\omega_m(t+\delta)$ separately in the two cavities with frequencies $\omega_0$ and $\omega'_0$.

Floquet theory is used to solve the Lindblad master equation, which directly relates $g^{(2)}_{\mathrm{HOM}}(0)$ to the spectral overlap between SPFCs \cite{supp}. When the modulation frequency is much larger than the dissipation rate and interaction strength ($g\ll \Omega$,  $\kappa+\gamma \ll \Omega$), the following condition holds
\begin{equation}\label{eq:g2final}
    g^{(2)}_\mathrm{HOM}(0) \approx
    \frac{1}{2}-
    \frac{1}{2}\left|\sum_n e^{-i\Omega\delta}s^*_ns_{n-1}\right|^2.
\end{equation}
A modulation signal satisfies Eq. (\ref{eq:cond})  is given by $\omega_m(t)=\frac{2A\Omega^2}{\pi^2}(\frac{\pi}{\Omega}-t)$, which follows a sawtooth waveform with a period of $T=\frac{2\pi}{\Omega}$ due to the nature of cavity modulators.

\begin{figure}[h]
    \centering
    \includegraphics[width=0.5\textwidth]{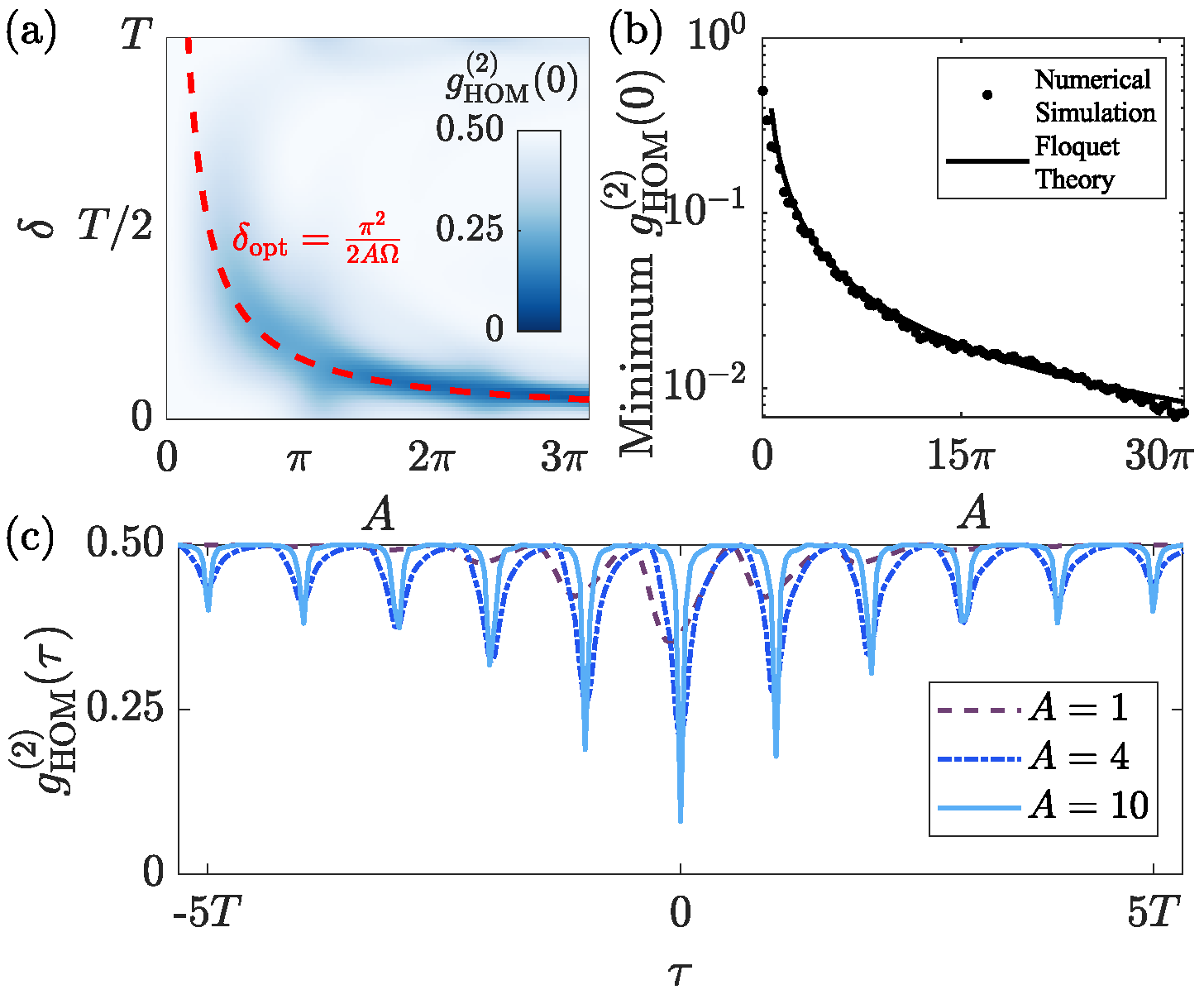}%
    \caption{(a) $g^{(2)}_\mathrm{HOM}(0)$ as a function of the modulation depth $A$ and signal time delay $\delta$. The red dashed line represents the optimal delay condition. (b) The optimal $g^{(2)}_\mathrm{HOM}(0)$ as a function of $A$. When $A$ increases, the number of overlapping sidebands for two SPFCs grows and the minimized $g^{(2)}_\mathrm{HOM}(0)$ gradually approaches 0. (c) The optimal $g^{(2)}_\mathrm{HOM}(\tau)$ curves at $A=1$, $4$, $10$, respectively. $g/(2\pi)=8$ GHz, $\gamma/(2\pi)=1$ GHz, $\kappa/(2\pi)=16$ GHz, and $\Omega/(2\pi)=50$ GHz are taken in the calculation \cite{englund2007controlling}. }
    \label{fig:result}
\end{figure}

Lindblad master equation is numerically solved to accurately model the time evolution of the density matrix and compute the second order correlation function as shown in Fig. \ref{fig:result}(a). With a modulation depth $A$, there is an optimal delay $\delta_{opt}=\frac{\pi^2}{2A\Omega}$ to minimize $g^{(2)}_\mathrm{HOM}(0)$.

Fig. \ref{fig:result}(b) shows the minimum values of $g^{(2)}_\mathrm{HOM}(0)$ as a function of the modulation depth $A$. The results obtained from the numerical calculation of Lindblad master equations (dotted line) match closely to those derived from Floquet theory (solid line), demonstrating that higher modulation depth enhances photon indistinguishability. Although numerical results exhibit a certain degree of characteristic fluctuations, our Floquet prediction yields a smoother curve because higher-order corrections are ignored in the simplified treatment \cite{hausinger2010dissipative, son2009floquet}. The findings enable a robust experimental design; for example, a cavity modulation depth exceeding $2.6\pi$ guarantees Hong-Ou-Mandel interference with $g^{(2)}_{\mathrm{HOM}}<0.1$, surpassing the fidelity required for mainstream QIP implementations \cite{aharonovich2016solid}. Using a pulse shaper to optimize the phase of each frequency component further reduces the required modulation depth $A$ \cite{supp} to $\pi$, a reachable level demonstrated in various frequency ranges \cite{yu2022integrated, zhou2024cavity, jin2014ultrafast}.

The correlation as a function of the time delay between single-photon arrivals $\tau$, is shown as $g^{(2)}_\mathrm{HOM}(\tau)$ in Fig. \ref{fig:result}(c). As the modulation depth increases, the narrowing of local event pulses within each temporal period is observed, leading to a high-contrast dip that enhances temporal resolution in the HOM experiment \cite{ren2014simultaneous}. This sharp dip and sufficient separation of pulses provide robust performance against incoherent temporal broadening, such as that introduced by group velocity dispersion and detector timing jitter \cite{yamazaki2023linear}. 

\begin{figure}[t]
    \centering
    \includegraphics[width=0.4\textwidth]{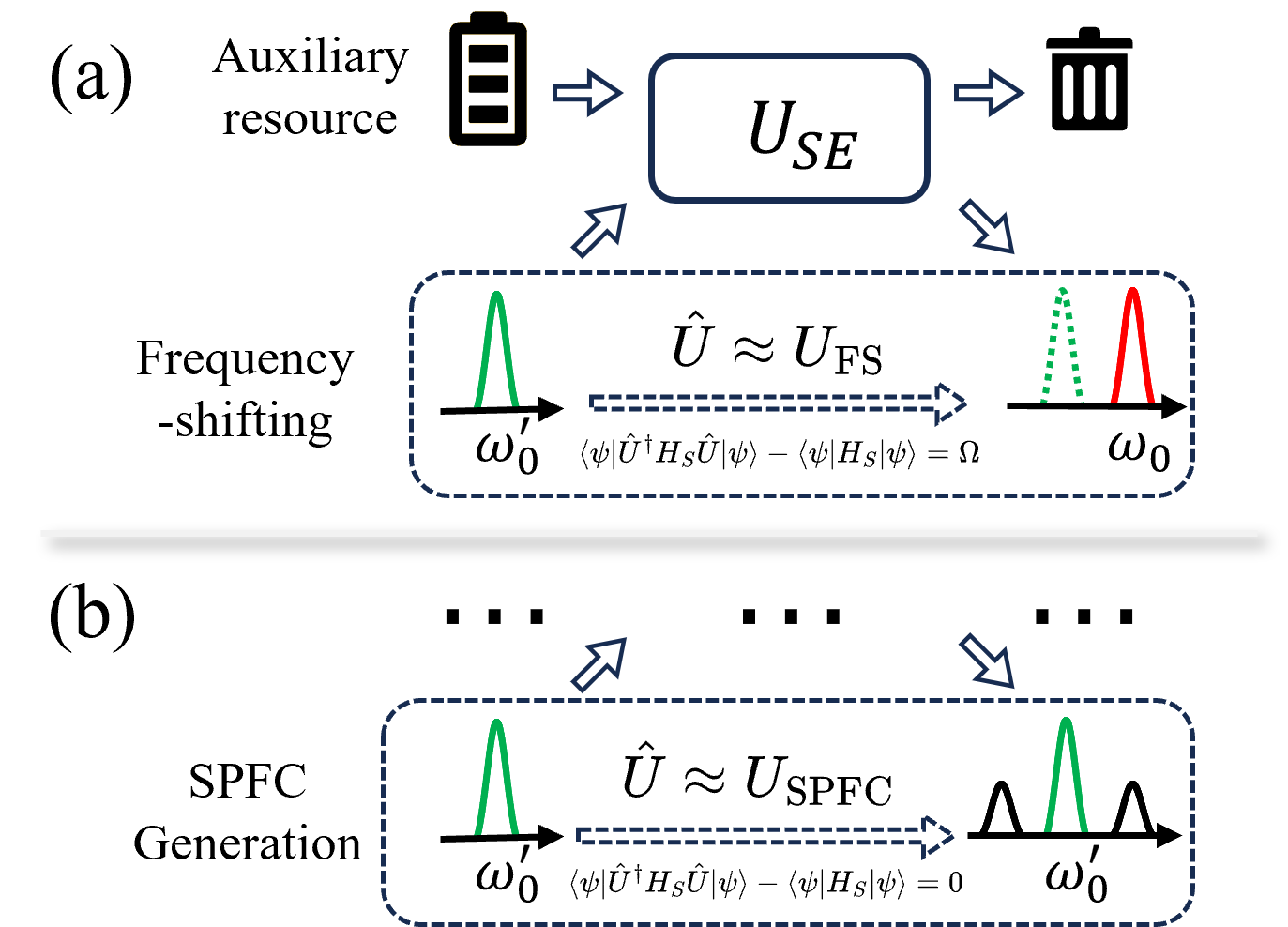}%
    \caption{A schematic of the microscopic models for frequency-shifting (a) and SPFC generation (b) from the perspective of quantum thermodynamics. The spectral manipulation of a single photon must involve an external auxiliary resource that supplies the necessary energy and other resources, thus together participating in a covariant operation $U_{\text{SE}}$. The processes $\hat U$ for system originating from $U_{\text{SE}}$ effectively describes the operation $U_{\text{FS}}$ for frequency-shifting and $U_{\text{SPFC}}$ for SPFC generation. In accordance with the laws of energy conservation, the external auxiliary resource must satisfy certain requirements to ensure the accuracy of energy-unconserving operations. Consequently, fundamental limits emerge, linking the operation accuracy, driving energy cost, and the energy change of the single photon state.}
    \label{fig:thermo}
\end{figure}

Our scheme maintains a low energy requirement under large spectral mismatches in contrast to conventional frequency‐shifting methods, which is confirmed by no-go theorems within the framework of quantum thermodynamics \cite{chiribella2021fundamental, yang2022energy}. As shown in Fig. \ref{fig:thermo}, the energy cost, which refers to the energy flows into / out of the single photon system and the extra necessary work done by the auxiliary system, should be considered as the driving energy in the experiment. Microscopically, the operation $\hat U$ on the single-photon state arises from the joint covariant evolution $U_{\text{SE}}$ of the photon and an external auxiliary system, followed by tracing out the auxiliary degrees of freedom. Trade-off relation appears in the microscopic model $U_{\text{SE}}$, as referred to the fundamental limitations in the theorems. They describe that a minimum energy cost is required to ensure the accuracy of a unitary operation. The generation of SPFCs, as we propose to achieve photon indistinguishability, is a unitary operation satisfying the law of energy conservation in the single-photon system. Therefore, the operation is free of these bounds and demonstrates significant energy-cost advantages over conventional frequency-shifting techniques. Specifically, to implement $U_{\text{SPFC}}$ by an evolution $\hat U$, the energy cost is not fundamentally bounded by the spectral mismatch $\Omega$, since the central frequency of the SPFC state remains unchanged during the generation process. In contrast, frequency-shifting operations $U_{\text{FS}}$ necessarily add or remove energy from single-photon states (an amount $\hbar\Omega$ for a shift by $\Omega$) \cite{kumar1990quantum, duguay1966optical}. The targeted unitary operation, $U_{\text{FS}}$, with high fidelity incurs an energy cost that is lower bounded by a quantity dependent on the spectral shift. It is worth noting that the theorems reveal that spectral shearing techniques incur an energy cost scaling of $\Omega^2$, while in our scheme the energy cost does not appear to scale with $\Omega$ in a straightforward manner. These facts are further confirmed by a classical optoelectronic model (see the supplementary material \cite{supp} for details) and experimental demonstrations \cite{fan2016integrated}.

In summary, the violation of energy conservation imposes trade-offs between the operational accuracy and spectral mismatch range with affordable energy cost. To overcome these limitations, we propose an energy-conserving operation to address spectral mismatch between distinct single photons. Our scheme enhances photon indistinguishability by increasing spectral overlap through SPFCs and frequency-driven modulation. The periodic time lens is utilized to generate flat-top SPFC states and a time delay to modulation signal is introduced to maintain the phase-locking condition. At a modulation depth of $A \approx 2.6\pi$, the second-order correlation function is suppressed to $g^{(2)}_{\mathrm{HOM}} = 0.1$, satisfying the fidelity requirements of most QIP applications~\cite{aharonovich2016solid}. If the pulse shaper is employed, despite its experimental complexity, the required modulation depth can be further reduced to $A\approx\pi$, which can be already satisfied in the current experimental condition \cite{yu2022integrated, zhou2024cavity, jin2014ultrafast}.

Furthermore, SPFC states are widely used in many frequency encoding schemes, which is advantageous for increasing the information density in quantum systems \cite{lukens2016frequency}. This strategy represents a new direction in achieving indistinguishability and is naturally compatible with spectral QIP schemes \cite{yamazaki2023linear, lukens2016frequency}, offering a promising alternative to conventional methods. Additionally, as spectral platforms continue to advance, it will be important to reconsider traditional challenges such as on-chip measurement and state generation, as these issues may be more readily addressed in spectral QIP systems compared to other schemes.

Besides, the generation of broad-spectrum, flat-top optical combs is also a significant focus in microwave photonics. Although the current study utilizes simple phase modulation, various alternative methods could efficiently implement the periodic time lens \cite{torres2008lossless, huh2015generation, lei2015observation, yu2022integrated, ma2024fast}. Recent advances have introduced new methods capable of producing flatter combs with lower power consumption \cite{dou2011improvement, wang2015optical, francis2020optical, crespo2023optimized}, offering more options to implement our photon-indistinguishability enhancement scheme.

\vspace{8pt}
\begin{acknowledgments}
This work is supported by the Zhejiang Province Leading Geese Plan (2024C01105), the National Natural Science Foundation of China (61574138, 61974131), and the China National Key Research and Development Program (2021YEB2800500).
\end{acknowledgments}

\bibliography{ref}

\end{document}